\documentclass[letterpaper]{aa}
\usepackage{graphicx}
\usepackage{txfonts}
\newcommand{\row}{eqnarray}
\newcommand{\calka}{\int^{\infty}_{0} d\nu^\prime \,}
\newcommand{\dnu}{\int^{\infty}_{0} d\nu \,}

\newcommand{\inu}{I_\nu}
\newcommand{\jp}{J_{\nu^\prime}}
\newcommand{\jnu}{J_\nu}

\newcommand{\bnu}{B_\nu}
\newcommand{\ptau}{d \tau_\nu}
\newcommand{\veps}{\varepsilon_\nu}
\newcommand{\naw}{(\nu,\nu^\prime)}
\titlerunning{Model atmospheres and X-ray spectra of iron-rich bursting 
    neutron stars}
\authorrunning{A. Majczyna et al.}

\begin{document}

\def\te{T_{e\hskip-0.5pt f\hskip-1.0pt f}}
\def\delt{\scriptstyle \Delta }

\title{Model atmospheres and X-ray spectra of bursting neutron stars}
\subtitle{II. Iron rich Comptonized Spectra} 

\author{A. Majczyna\inst{1}, J. Madej\inst{2}, P.C. Joss\inst{3}, 
A. R\'o\.za\'nska\inst{1}}
\offprints{A. Majczyna \\ \email{majczyna@camk.edu.pl}}
\institute {$^1$ Copernicus Astronomical Center, Bartycka 18, PL-00-716 Warsaw,
Poland\\
$^2$ Warsaw University Observatory, Al. Ujazdowskie 4, 00-478 Warsaw, Poland \\
$^3$ Massachusetts Institute of Technology, Cambridge, MA 02139, U.S.A.}
\date{Received ...; accepted ...}

\abstract{
This paper presents the set of plane-parallel model atmosphere equations
for a very hot neutron star (X-ray burst source). The model equations assume
both hydrostatic and radiative equilibrium, and the equation of state of
an ideal gas in local thermodynamic equilibrium (LTE). The equation of
radiative transfer includes terms describing Compton scattering of photons
on free electrons in fully relativistic thermal motion, for photon
energies approaching $m_e \, c^2$. Model equations take into account many
bound-free and free-free energy-dependent opacities
of hydrogen, helium, and the iron ions, and also a dozen bound-bound 
opacities for the highest ions of iron. We solve model equations by
partial linearisation and the technique of variable Eddington factors.
Large grid of H-He-Fe model atmospheres of X-ray burst sources has been
computed for $10^7 \le \te \le 3\times 10^7$ K, a wide range of surface
gravity, and various iron abundances.  
We demonstrate that the spectra of X-ray bursters with iron present in
the accreting matter differ significantly from pure H-He spectra 
(published in an earlier paper), and also from blackbody spectra. 
Comptonized spectra with significant iron abundance are generally closer
to blackbody spectra than spectra of H-He atmospheres. 
The ratio of color to effective temperatures in our grid always remains in 
the range $1.2 < T_c/\te < 1.85$. 
The present grid of model atmospheres and theoretical X-ray spectra will
be used to determine the effective temperatures, radii and $M/R$ ratios of
bursting neutron stars from observational data.

   \keywords{Radiative transfer -- Stars: atmospheres -- Stars: neutron
             -- X-rays: bursts -- X-rays: spectroscopy   }

}

\maketitle
\section{Introduction}

X-ray bursters were discovered by Grindlay et al. (1976), and
Belian et al. (1976). These X-ray sources are neutron stars in interacting
binaries. Very small mass functions and short orbital periods indicate
that the secondary star has low mass (Stella et al. 1987\cite{stella};
Chakrabarty \& Morgan 1998). Periodic pulsations have not been detected in
the light curves of the great majority of X-ray burst sources, and this has
been used as an argument that the surface magnetic field of a neutron star
is weak (Joss \& Li 1980)\footnote{Pulsations were observed in 
SAX J1808.4-3658, but 
other features (e.g. kHz QPO) suggested that the magnetic field is 
relatively weak. Recently at least 4 accreting millisecond pulsars have 
been detected, and all are objects with weak magnetic fields. }. 
We note that the spectrum of a type I X-ray burst becomes softer during the
decay of the burst, and that such a softening is the signature of this 
type of events. 

X-ray bursts are very energetic events. The estimated energy released 
during a single burst is typically $\sim 10^{39}$ ergs. The light curve
of a burst is characterized by a rise time of $\sim$ 1s and a decay time of 
$\sim$ 3-100s. X-ray bursts from a given source are recurrent events, but are
not strictly periodic. The intervals between bursts are typically in the range 
of $\sim 10^4-10^5$s, but sometimes a burster undergoes inactive phases lasting
weeks or even months (see review articles by Joss \& Rappaport 1984; 
Lewin, van Paradijs \& Taam 1993). 

The origin of an X-ray burst is a thermonuclear flash in freshly accreted
matter on the neutron-star surface. This idea was proposed by Woosley
\& Taam (1976) and Maraschi \& Cavaliere (1977). This model explains quite
well many of the global features of X-ray bursters, including 
energetics, temporal structure and spectral behavior of an average X-ray
burst source (Joss 1977, 1978; Ayasli \& Joss 1982).

Neutron star spectra were initially assumed to be perfect blackbodies.
This assumption, however, is not valid for any realistic stellar
atmosphere. For the case at hand, significant distortions from a simple
blackbody spectrum result from radiative transfer through the neutron-star
atmosphere, because the radiative opacity is dominated 
by scattering on free electrons (Madej 1974; Czerny \& Sztajno 1983). In 
this case the outgoing X-ray spectra are shifted towards higher
energies and the peak X-ray flux is shifted in the same direction both
for coherent (Thomson) and incoherent (Compton) scattering. 
There are two reasons for this effect. First, the free-free 
absorption coefficient in the atmosphere of a hot neutron star varies
with frequency as $\nu^{-3}$ (for the opacity by heavier elements this 
relation is flat), and this causes a redistribution of radiation 
towards higher energies. Second, the presence of photon scattering causes an 
additional shift of the outgoing flux in the same direction. The latter 
effect results from the shift of the thermalization depth to deeper
(hotter) layers in the atmosphere. In other words, in the scattering 
atmosphere of a neutron star, photons are created by thermal processes
below the photosphere, where the local temperature $T$ is much higher than
the effective temperature $\te$, and where the peak of the emission
spectrum is shifted to higher energies. 

An exact description of Compton scattering (Sampson 1959; Pomraning 1973; 
Madej 1989, 1991a; Psaltis \& Lamb 1997; Psaltis 2001) is very important for
the accurate determination of basic neutron star parameters in X-ray bursters
by use of fitting procedures. For this reason many authors have
computed models of neutron star atmospheres with various degrees of
sophistication in the treatment of Compton scattering.
The influence of Compton scattering on the atmospheres of X-ray bursters 
has been discussed in several papers (see e.g. Ebisuzaki, Hanawa \&
Sugimoto 1984; and Foster, Ross \& Fabian 1986). Consistent models of the
atmosphere with Compton scattering have been presented by London, 
Taam \& Howard (1986), Ebisuzaki (1987), and Ross \& Fabian (1993). 
In these papers the authors utilised the Kompaneets approximation
for the description of Compton scattering. 

We also note the works by Babul \& Paczy\'nski (1987),
Joss \& Melia (1987), Zavlin \& Shibanov (1991a,b)
and Pavlov et al. (1991), who computed semianalytic model atmospheres and
X-ray spectra of bursting neutron stars with luminosities approaching the
Eddington limit. Most of these papers predicted the basic qualitative
features of Comptonised model atmospheres and spectra, such as the
shift between color and effective (or blackbody) temperatures. Another
important feature found by these authors is the existence of a flux
excess at low X-ray energies in the Comptonized emergent spectra
of models with luminosities approaching the Eddington limit. 

Recently, papers with more general descriptions of Compton 
scattering were published by Madej (1991a,b), Madej \& R\'o\.za\'nska 
(2000a,b), Joss \& Madej (2001), and Madej, Joss \& R\'o\.za\'nska (2004).
The equation of transfer assumed in the latter papers allows for a large
energy exchange beetwen photons and electrons, and therefore enables one
to compute Compton scattering of photons with large initial energies,
as well as the corresponding redistribution of photon energies during 
scattering. However, in these works the linear polarization of 
radiation was not considered, and the differential Compton scattering 
coefficient was averaged over scattering angles. 

The present paper presents model atmospheres and spectra that will be
used in the future for the quantitative interpretation of the observed X-ray
spectra of bursting neutron stars. This paper is organised in the following
manner: Section 2 presents the basic assumptions and equations of the model
atmosphere with Compton scattering taken into account. Section 3 
presents our computations of model atmospheres containing iron and 
the resultant theoretical X-ray spectra, and compares the 
results with the H-He models of our previous paper (Madej,
Joss \& R\'o\.za\'nska 2004). Section 4 presents a summary and conclusions. 

\section{Our models - assumptions and equations}
\subsection{Assumptions}
We formulate the set of model neutron-star atmosphere equations assuming:
\begin{itemize} 
\item plane-parallel geometry for the atmosphere,
\item local thermodynamical equilibrium and the equation of state of
   an ideal gas,
\item radiative and hydrostatic equilibrium, 
\item zero magnetic field,
\item no neutron-star rotation.
\end{itemize}

The equation of transfer includes free-free and bound-free absorption from all 
ions of hydrogen, helium and iron, as well as Compton scattering terms. 
The latter allow us to trace scattering off electrons with relativistic
thermal motion, and with initial photon energies exceeding the
electron rest mass. Moreover, the opacity of 10 spectral lines of highly
ionised iron was included in our calculations. All these lines belong to
the fundamental series of helium-like and hydrogen-like iron, and they fall
in the energy range 6.7 - 9.0 keV, where most of the radiative energy
of an X-ray burster is emitted. The total number of considered lines was
chosen arbitrarily. 

We are aware that the assumed chemical composition probably does not 
correspond to the atmosphere of a real X-ray burster. The composition
of both the freshly accreted matter and the material processed
during thermonuclear flashes in the stellar envelope will be 
rich of CNO and a variety of heavier elements. In our
models, however, iron represents the average ``metal'', serving as
the contributor of free electrons and the strong additional
continuum free-free opacities. In this way we investigate the influence
of both of these agents on the model atmospheres and the Comptonisation
of the outgoing spectra, no matter what specific heavy elements may be
the source of these effects in a real X-ray burster. We are then
able to compare the resultant spectra with the
spectra of pure H-He atmospheres (Madej, Joss \& R\'o\.za\'nska 2004).

Ionization equilibrium in the model was determined in the following
manner. For each discrete standard optical depth level $\tau_i$ and
the corresponding values of temperature $T$ and gas pressure $P_g$, we 
have guessed the value of electron pressure $P_e$ and then determined
trial populations of all possible ionization states by solving the set
of Saha equations for each element. Ionization populations computed
in this way determine new values of electron density $N_e$, 
electron pressure $P_e$, and the resulting gas pressure $P_g^{st} =
P_g^{st} (P_e,T) = (N_{at} + N_e) \, kT$ (ideal gas approximation).
This procedure was iterated until $P_g^{st}$ obtained in the above manner
approached $P_g$ with a relative accuracy better than $10^{-5}$.

The energy-dependent absorption $\kappa_\nu$ in our model
atmosphere code, ATM21, is the sum of many agents. First, we always 
compute bound-free opacities from the 9 lowest levels of hydrogen, the 30
lowest levels of neutral helium, and the 10 lowest levels of
singly ionized helium. Second, in this work bound-free opacities of
all existing iron ions from the ground level were computed following
Verner \& Yakovlev (1995), who have published formulae fitting 
opacities of arbitrary elements and ions with atomic number $Z\le 30$,
stored in the Opacity Project database, see Seaton (1987). Third, free-free 
opacities of all existing ions were computed using the standard equations
from Mihalas (1978).

General relativistic effects, such as the gravitational redshift of
radiation and the bending of photon trajectories in strong gravitional field
were not included in our equations. See
Madej, Joss \& R\'o\.za\'nska (2004) for a discussion of some of the
effects of general relativity on the emergent spectra.

The atmosphere of a neutron star with $\te \ge 10^7$ K is  
geometrically very thin, usually of the order of 1-10 m (see Madej 1991b),
except cases of the lowest values, $g$, for gravitational
acceleration at the neutron star surface, when the
model atmosphere approaches the critical gravity $g_{cr}$. 
As $g$ approaches $g_{cr}$, the acceleration exerted on the atmosphere
by the radiation field balances the downward gravitational
acceleration, and the atmosphere expands to infinity.
For higher values of $g$, the ratio of the height of the atmosphere
to the neutron- star radius is much smaller than unity. In this
case the assumption of plane-parallel structure for the atmosphere is
fully acceptable. The existence of radiative equilibrium implies
that the radiation field is the only means of energy
transport, and both convection and thermal conductivity can be neglected. 
Magnetic fields in X-ray bursts sources are relatively weak, and are in 
the range $10^7-10^9$ G (Joss \& Li (1980); Lewin, van Paradijs \&
Taam 1995; Miller, Lamb \& Psaltis 1998). Consequently, the
magnetic field is unable to modify the continuum and line opacities.

Our model equations describe the transfer of radiation subject to the 
constraints of hydrostatic and radiative equilibrium.
We assume the equation of state of an ideal gas in local thermodynamic
equilibrium (LTE), so that all occupation numbers of bound and free 
states, opacities and emissivities are the same as their 
thermal equilibrium values at the local temperature $T$ and density
$\rho$ throughout the entire atmosphere. 
However, in our model atmospheres the equation of transfer
is dominated by the Compton scattering terms. This means explicitly
that the equation of transfer distinctly remains in non-LTE.

\subsection{Model equations}
The equation of transfer with Compton scattering terms is as follows: 
\begin{\row}
&&\mu \, \frac{\partial I_{\nu}(z,\mu)}{\rho \, \partial 
z}=\kappa^{\prime}_{\nu}\, (1-e^{-h\nu/kT}) \, (B_\nu 
-I_\nu) \\ \nonumber
&+&(1+\frac{c^2}{2h\nu^3}I_\nu)\oint_{\omega^{\prime}}
\frac{d\omega^\prime}{4\pi} \int^{\infty}_{0} d\nu^{\prime} 
\frac{\nu}{\nu^\prime} \, \sigma(\nu^\prime \to \nu, \vec{n}\cdot 
\vec{n^\prime}) \, I_{\nu^\prime} \\ \nonumber
&-&I_\nu \oint_{\omega^{\prime}}\frac{d\omega^\prime}{4\pi} 
\int^{\infty}_{0} d\nu^{\prime} \sigma(\nu \to \nu^\prime, \vec{n}\cdot 
\vec{n^\prime }) \, (1+\frac{c^2}{2h\nu^{\prime 3}}\, I_{\nu^{\prime}})\, ,
\label{eqn:rtrans}
\end{\row}
where $B_{\nu}$ is the Planck function, and $\inu$ and $J_{\nu}$ denote 
specific intensity and mean intensity of radiation, respectively. Parameter
$\kappa^\prime_\nu$ denotes the true absorption coefficient without 
correction for induced emission, whereas 
$\kappa_\nu = \kappa^\prime_\nu \, (1-e^{-h\nu/kT}) $
will be used later as the same coefficient corrected for induced emission.
Variable $z$ denotes geometrical depth, and $\mu$
is the cosine of the zenithal angle (Madej 1991a). The variable 
\begin{equation}
\sigma(\nu \to \nu^\prime, \vec{n} \cdot \vec{n^\prime })=\sigma_\nu \,
\phi(\nu,\nu^{\prime},\vec{n} \cdot \vec{n^\prime })
\end{equation}
denotes the differential cross section of Compton scattering, where the
redistribution function $\phi(\nu,\nu^{\prime},\vec{n}\cdot \vec{n^\prime})$
is normalised to unity:
\begin{equation}
\oint_{\omega^{\prime}}
\frac{d\omega^\prime}{4\pi} \int^{\infty}_{0} d\nu^{\prime} 
\phi(\nu,\nu^{\prime},\vec{n}\cdot \vec{n^\prime})\,= \,1 \, .
\end{equation}
There exists the universal relation
\begin{equation}
\sigma ( \nu \rightarrow \nu {^\prime }, \vec n \cdot \vec n {^\prime} )
\, \nu ^2  e^{-h\nu /kT} = \,
\sigma ( \nu {^\prime }\rightarrow \nu , \vec n {^\prime} \cdot \vec n )
\, {\nu ^\prime }^2 e^{-h\nu ^\prime /kT}  \> ,
\label{equ:sig}
\end{equation}
which results from the detailed balancing of Compton scattering in 
thermodynamic equilibrium (cf. Pomraning 1973).

The differential cross section $\sigma(\nu \to \nu^\prime, \vec{n} \cdot
\vec{n^\prime })$ has been computed here following Guilbert (1981). 
Unfortunately, the set of the corresponding equations is very complex and 
cannot be presented explicitly here. 

We can integrate $\phi(\nu,\nu^{\prime},\vec{n}\cdot \vec{n^\prime})$ over
the solid angles $\omega^\prime$, and obtain the zeroth moment of the 
redistribution function:
\begin{equation}
\Phi\naw = \oint \frac{d\omega^\prime}{4\pi} \,
\phi (\nu,\nu^\prime,\vec{n}\cdot \vec{n^\prime}) \, .
\end{equation}
The equation of radiative transfer, Eq. 1, can be transformed to a
useful form using the following approximations. First, we replace the
redistribution function $\phi(\nu,\nu^{\prime},\vec{n}\cdot
\vec{n^\prime})$
by its zeroth moment $\Phi\naw$. Second, the specific intensity of radiation
$I_\nu$ in both stimulated scattering terms was replaced by the mean 
intensity $J_\nu$. Then, we
define two new Compton redistribution functions:
\begin{equation}
\Phi_1 (\nu,\nu^\prime)=\left(1+\frac{c^2}{2h\nu^{\prime \, 3}} \jp 
\right)\, \Phi(\nu,\nu^\prime) \, ,
\end{equation}
\begin{equation}
\Phi_2 
(\nu,\nu^\prime)=\left(1+\frac{c^2}{2h\nu^3}\jnu\right) \,
\left(\frac{\nu}{\nu^\prime}\right)^3 \exp\left[-\, 
\frac{h(\nu-\nu^\prime)}{kT}\right] \, \Phi(\nu,\nu^\prime) \, . 
\end{equation} 
Third, we arbitrarily assume that the angular integral of 
$\sigma(\nu^\prime \to \nu, \vec{n}\cdot \vec{n^\prime}) \, I_{\nu^\prime}$
in Eq. 1 can be approximated by the product of two corresponding angular
integrals of $\sigma(\nu^\prime \to \nu, \vec{n}\cdot \vec{n^\prime})$ and
$I_{\nu^\prime}$. Eq. 1 changes to
\begin{eqnarray}
\label{trans2}
&& \mu\, \frac{\partial
J_{\nu}}{\rho \, \partial z}=\kappa_{\nu} \, (B_{\nu}- I_{\nu}) 
- \sigma_\nu \, I_{\nu}\int_0^{\infty} 
\Phi_1(\nu,\nu') \, d\nu' \\
&& \hskip12mm + \, \sigma_\nu \int_0 ^{\infty}
J_{\nu'}\, \Phi_2(\nu,\nu') \, d\nu'  \nonumber
\end{eqnarray} 

The equation of transfer can be written on the monochromatic optical depth
scale, $d\tau_\nu=- \, (\kappa_\nu +\sigma_\nu)\, \rho \, dz \, $. 
After simple algebra we obtain
\begin{eqnarray}
\label{transfer}
&& \mu\, \frac{\partial
I_{\nu}}{\partial\tau_{\nu}}=I_{\nu}-\veps \,B_{\nu}-(1-\veps)\, J_{\nu}+{} \\
&&{}(1-\veps)\, J_{\nu}\int_0^{\infty}
\Phi_1(\nu,\nu') \, d\nu'-(1-\veps)\int_0 ^{\infty}
J_{\nu'}\, \Phi_2(\nu,\nu') \, d\nu' \, ,  \nonumber
\end{eqnarray}
where $\veps = \kappa_\nu / (\kappa_\nu + \sigma_\nu) $ is the ratio of 
true absorption to the total opacity coefficients, and Compton scattering
is described by the redistribution functions $\Phi_1(\nu,\nu')$ and 
$\Phi_2(\nu,\nu')$.

Following the standard approach we obtain the zeroth and first moments of 
the equation of transfer in the following form:  
\begin{\row}
&&\frac{d K_\nu}{\ptau}=H_\nu \\ \nonumber
\label{row4}
&&\frac{d H_\nu}{\ptau}=\veps(\jnu-\bnu) +(1-\veps) \, \jnu \calka 
\Phi_1\naw \\ 
&&\hskip10mm - \, (1-\veps) \calka \Phi_2\naw \, \jp 
\end{\row}
Combining both the above equations we obtain the second order differential
equation of transfer
\begin{\row}
\eta_\nu \frac{d}{d 
\tau}\left(\eta_\nu \frac{d}{d \tau}(f_\nu\jnu)\right)& =&
\veps(\jnu-\bnu^*) \\ \nonumber
&+& (1-\veps)\, \jnu\calka \Phi_1^*\naw \\ \nonumber
&-& (1-\veps)\calka \Phi_2^*\naw \, \jp 
\end{\row}
where $\eta_\nu=(\kappa_\nu +\sigma_\nu )/(\kappa +\sigma)_{std}$, 
$f_\nu=K_\nu/\jnu$ is the variable Eddington factor, and $\tau$ denotes the
standard optical depth. The latter is the monochromatic
optical depth computed at the fixed (standard) wavelength. 
 
Terms with an asterisk are calculated at the unknown temperature at which the
equation of radiative equilibrium is precisely fulfiled. Since we seek 
this temperature, we make a linearization of the three following functions,
expanding them in a Taylor series with respect to temperature
\begin{\row}
&&\bnu^\ast(T) =\bnu(T)+\Delta T \left(\frac{\partial \bnu}{\partial T}\right)_\tau \, , \\
&&\Phi_1^\ast\naw = \Phi_1\naw +\Delta T \left(\frac{\partial \Phi_1}{\partial
T}\right)_\tau \, , \\
&&\Phi_2^\ast\naw = \Phi_2\naw +\Delta T \left(\frac{\partial \Phi_2}{\partial
T}\right)_\tau \, , 
\end{\row}
where $\Delta T= T^* - T$. The second and higher order derivatives were 
neglected in the above expansions.

\subsection{Radiative equilibrium}
The equation of radiative equilibrium requires that
\begin{equation}
\int^\infty _0  H_\nu \, d\nu = \frac{\sigma_R \te^4}{4\pi}
\end{equation}
Differentiating the above equation with respect to $\tau $, and using
Eq.~\ref{row4} yields the useful equation of radiative equilibrium
\begin{\row}
\label{rownowaga}
0&=& \dnu \, \eta_\nu \veps \,(\jnu-\bnu^{\ast}) \\ \nonumber
&+&\dnu\, (1-\veps)\jnu \calka \Phi_1^{\ast}\naw \\ \nonumber
&-&\dnu\, \eta_\nu (1-\veps)\calka \Phi_2^{\ast}\naw \jp
\end{\row}
The above equation is linearised according to the procedure
described in Section 2.2. After that we obtain the temperature 
corrections in the form:
\begin{\row}
\Delta T=\frac{\dnu \,\veps \eta_\nu\, (\jnu-\bnu)+L(\tau)}{\dnu\, 
\veps\eta_\nu \left(\partial \bnu / \partial T\right) -L'(\tau)}
\end{\row}
where functions $L(\tau)$ and $L'(\tau)$ are defined as follows:
\begin{\row}
L(\tau)&=&\dnu (\jnu-\bnu) \calka \Phi_1\naw \times \\ \nonumber
&&\hskip1.5cm \left[ \eta_\nu (1-\veps)
  -\eta_{\nu^\prime}
  (1-\varepsilon_{\nu^\prime})\left(\frac{\nu^\prime}{\nu}\right)\right] 
\end{\row}
\begin{\row}
L'(\tau)&=&\dnu \left[\eta_\nu (1-\veps)\jnu \calka \frac{\partial
\Phi_1}{\partial T} \right. \\ \nonumber 
&& \hskip1.5cm \left. - \eta_\nu (1-\veps)\calka \jp \frac{\partial
\Phi_2}{\partial T}\right] 
\end{\row} 

\subsection{Hydrostatic equilibrium}
Our model assumes also the equation of hydrostatic equilibrium, since we
investigate static atmospheres. The  condition of hydrostatic equilibrium
can be written in the form:
\begin{equation}
\frac{dP}{dz}=\frac{dP_g}{dz}+\frac{dP_r}{dz}=- \, g\rho \, .
\end{equation}
Here the total scalar pressure $P$ is the sum of gas and radiation 
pressures, $P=P_g+P_r \,$.
If we write the above equation on the standard optical depth scale $\tau$,
and use the expression
\begin{equation}
\frac{dP_r}{d\tau}=\frac{4\pi}{c}\int ^{\infty}_0 \eta_\nu H_\nu\, d\nu \, ,
\end{equation}
then we get the final form of the hydrostatic equlibrium equation:
\begin{equation}
\frac{dP_g}{d\tau}=\frac{g}{(\kappa+\sigma)_{std}}-\frac{4\pi}{c}
\int^{\infty}_0 \eta_\nu H_\nu \, d\nu\, .
\end{equation}

\subsection{The equation of transfer and boundary conditions}
The final equation of transfer, which is adopted in numerical calculations,
is of the form:
\begin{eqnarray}
&&\eta_\nu \, \frac{\partial}{\partial 
\tau}\left(\eta_\nu \frac{\partial}{\partial\tau_\nu}(f_\nu\jnu)\right) 
=\epsilon_{\nu}\,(J_{\nu}-B_{\nu}){}\nonumber \\
&&{}+(1-\epsilon_{\nu}) \, J_{\nu}\int_0^{\infty}
\Phi_1(\nu,\nu')\, d\nu'-(1-\epsilon_{\nu})\int_0^{\infty}
J_{\nu'}\, \Phi_2(\nu,\nu'){}
\nonumber \\
&&{}-\Big[\epsilon_{\nu}\frac{\partial
B_{\nu}}{\partial T}-(1-\epsilon_{\nu})\, J_{\nu}\int_0^{\infty}
\frac{\partial \Phi_1}{\partial T}\, d\nu'+(1-\epsilon_{\nu})\int_0^{\infty}
J_{\nu}\, \frac{\partial \Phi_2}{\partial T}\, d\nu'\Big] {}
\nonumber \\
&&{}\times
\frac{\int_0^{\infty}\eta_{\nu}\epsilon_{\nu}\,(J_{\nu}-B_{\nu})\,d\nu+L(\tau)}
{\int_0^{\infty}\eta_{\nu}\epsilon_{\nu}\, (\partial B/\partial
T) \, d\nu -L'(\tau)}
\end{eqnarray} 
The equation can be solved when both the upper and the lower boundary 
conditions are specified.  \\

The upper boundary condition is of the standard form,
\begin{equation}
\frac{\partial}{\partial \tau_\nu} (f_\nu \jnu) = h_\nu \jnu(0)
\end{equation}
where $h_\nu=H_\nu(0)/\jnu(0)$ is the surface scalar factor which has to
be iterated simultaneusly with the variable Eddington factors $f_\nu$
(Mihalas 1978).   \\
 
The lower boundary condition is in fact the thermalisation condition:
\begin{equation}
\frac{\partial}{\partial \tau_\nu}\left(f_\nu \jnu \right)=\frac{1}{3}\, \frac{\partial
  \bnu}{\partial \tau_\nu}
\end{equation}
A detailed description and transformations of the inner boundary
condition to the useful form are given in the Appendix.

\section{Results}

We calculated a grid of 106 models. Each of them was computed on a mesh
of 144 standard optical depth points (12 points per decade), distributed
from $\tau_{std}=10^{-8}$ to $10^4$, and on 1139 frequency (energy) 
points, ranging from 400 keV to 12 eV. Effective temperatures of our
neutron star atmospheres are $\te$=1, 1.5, 2, 2.5, 3$\times10^7$
K, and surface gravities $g$ range from the critical gravity $g_{cr}$ 
to $10^{15}$ (cgs units). Chemical composition includes H and He in the
solar abundance $N_{He}/N_H=0.11$, and iron 
either $N_{Fe}/N_H=3.7\times10^{-5}$ (solar value), or 100 times higher. 
The constancy of bolometric flux across each model atmosphere was better than
0.1\%, and only at a few points located deep below the photosphere
was the flux error worse, up to about 2.5\%.

\subsection{Temperature structure of models}

The distribution of temperature in a model atmosphere in radiative
equilibrium is determined by several different parameters, such as the 
effective temperature, surface gravity, and the particular chemical 
composition. Assumption of either Compton or Thomson scattering also is of
crucial importance. In order to demonstrate particular features in $T(\tau)$ 
stratification we have arbitrarily chosen models with the effective 
temperature $\te=2\times 10^{7}$. In all such models, the gradient of 
temperature is very steep in deep optical depths. In higher layers, e.g. for
$\tau_{Ross} < 0.1$, the run of $T(\tau_{Ross})$ exhibits different
behaviour.

In the case of models with Thomson scattering, temperature decreases with 
decreasing $\tau_{Ross}$ to some minimum value, and above this minimum 
$T$ can rise only slightly and the model atmosphere is
isothermal. A model atmosphere can be isothermal starting from layers
where it is optically thin, and photons do not efficiently
interact with gas. In the case of a model atmosphere with Compton scattering
the situation is qualitatively different, since Compton scattering allows for
energy and momentum exchange between photons and free electrons. 
This process causes heating of the scattering layers of the atmosphere,
because energy is transfered from radiation created in a hot layer of 
thermalisation to colder electron gas near the surface. Such an effect 
explains the existence of a temperature rise in the outermost layers,
where Compton scattering dominates and sources of absorption vanish due
to extremely low density and complete ionisation of atoms.

\begin{figure}[!h]
\resizebox{0.95\hsize}{7cm}{\rotatebox{0}{\includegraphics[scale=0.4]{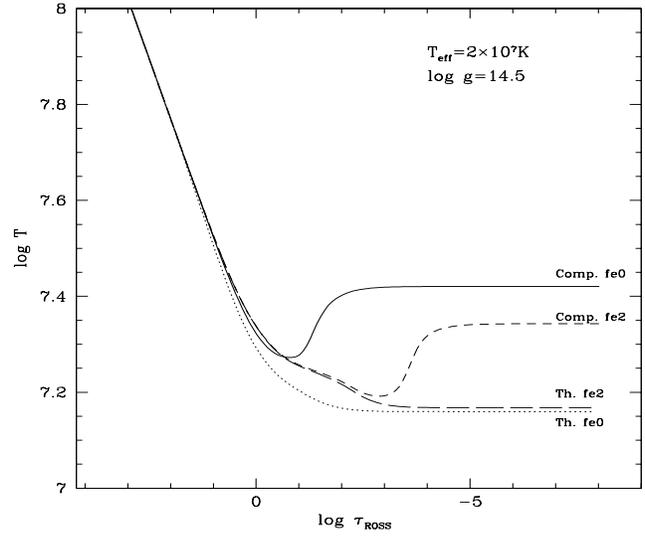}}}
\caption{\small{Temperature distribution in atmospheres of $\te=2\times
10^7$ K  and $\log{g}=14.5$, and various chemical compositions. Both 
Compton and Thomson scattering models are shown here. The solid line 
represents an atmosphere with solar iron abundance $N_{Fe}/N_H =
3.7 \times 10^{-5}$ and the dashed line represents a model with the iron
abundance 100 times higher (Compton scattering models). Dotted
line: model with solar iron abundance, long dashed line -- 100 times
higher iron abundance (Thomson scattering). }}
\label{f1}
\end{figure}

On the contrary, a model atmosphere computed under an assumption of coherent
Thomson scattering exhibits a monotonic run of temperature $T$, decreasing
outwards. The isothermal outermost region spreads above some optical depth 
located higher than in the Compton scattering model (see Fig. \ref{f1}).
Moreover, the isothermal zone exhibits a much lower temperature. This is
because Thomson scattering implies zero energy exchange between photons 
and electrons, consequently, it cannot extract thermal energy from hard
radiation created in hot thermalisation layers. In particular, in the 
model of $\te=2\times 10^7$ K, $\log{g}=14.5$, and the solar iron 
abundance, the atmosphere becomes isothermal at the Rosseland optical depth 
$\tau_{Ross}\approx 0.3$ for Compton scattering and 0.03 for Thomson 
scattering atmospheres. Boundary temperatures in both cases are
$T_0 \approx 2.21 \times 10^7$ K and $1.45 \times 10^7$ K, respectively.

\begin{figure}[!ht]
\resizebox{0.95\hsize}{7cm}{\rotatebox{0}{\includegraphics[scale=0.4]{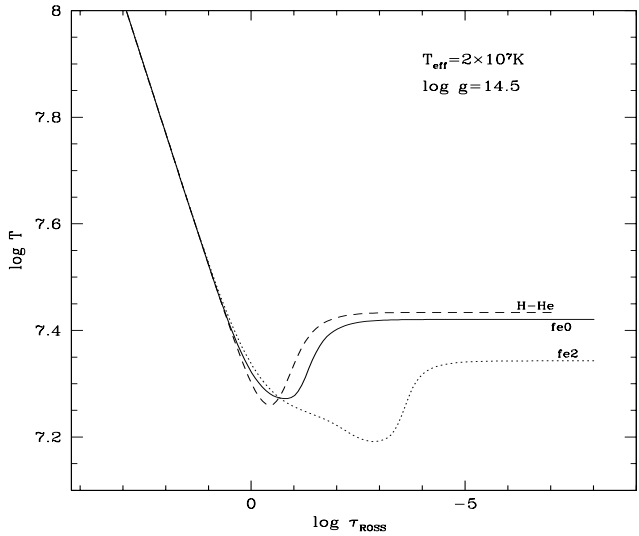}}}
\caption{\small{Temperature stratification in atmospheres of $\te=2\times
10^7$ K and $\log{g}=14.5$, and different iron abundances.
Solid line -- solar abundance of iron, $N_{Fe}/N_H=3.7 \times 10^{-5}$,
dotted line -- iron abundance is 100 times higher; dashed line 
-- H-He atmosphere.  }}
\label{f2}
\end{figure}

In a model atmosphere with nonzero iron abundance, thermal absorption is
higher than in a H-He atmosphere of the same $\te$ and $\log g$. This
is because iron is a strong opacity source. In particular, iron adds
huge free-free absorption to the model atmosphere. Such an atmosphere is
cooler in its uppermost layers, see Fig.~\ref{f2}. This is because thermal
nongrey absorption in LTE is a very efficient cooling mechanism due 
to the Kirchoff law, cf. also Mihalas (1978). 
Thermal emissivity and absorption in LTE fulfil the equation
$j_{\nu}=\kappa_{\nu}B_{\nu}$, 
where $j_{\nu}$ denotes emissivity, $\kappa_{\nu}$ is the thermal absorption 
coefficient, and $B_{\nu}$ denotes the Planck function. 
This relation implies that with increasing absorption, the rate of 
emission of radiation increases. If there exists a spectral window where
the optical depth is low, then this layer easily loses thermal energy
and its temperature $T$ decreases. An atmosphere gets isothermal starting
from the layer in which photons very weakly interact with matter. 

\begin{figure}[!ht]
\resizebox{0.95\hsize}{7cm}{\rotatebox{0}{\includegraphics[scale=0.4]{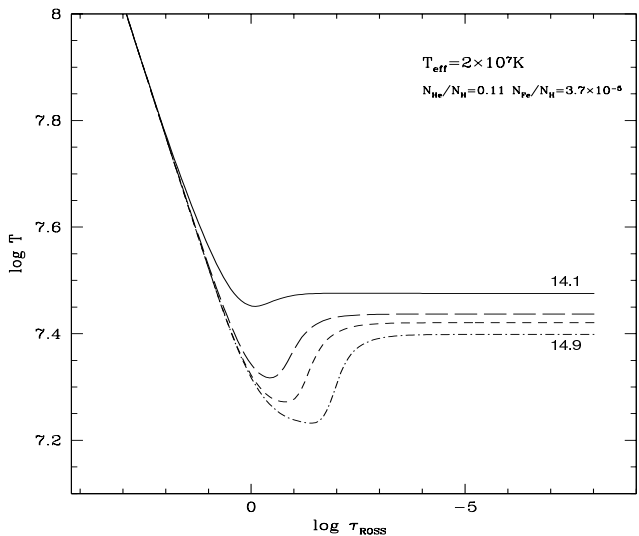}}}
\caption{\small{Temperature structure of atmospheres at the fixed 
$\te=2\times 10^7$ K and solar iron abundance, and different surface
gravities. Solid line -- $\log{g}=14.1$, dashed line -- 14.3, dotted
line -- 14.5, dot-dashed line -- 14.9.}}
\label{f3}
\end{figure}

Surface gravity has an essential impact on the structure of an atmosphere
(see Fig.~\ref{f3}). In the case of high surface gravity, the density
and thermal absorption coefficient $\kappa_\nu$ also are high. Consequently,
photons are more efficiently absorbed and there exists a deeper minimum of
temperature $T$ than in the atmosphere of a neutron star with lower gravity.
In a model with low surface gravity, the temperature $T$ in the outermost
layers is higher than in a model with high gravity, since Compton scattering
is the dominant opacity source and gas is effectively heated by 
multiple photon-electron energy exchanges.

\begin{figure}[!h]
\resizebox{0.95\hsize}{7cm}{\rotatebox{0}{\includegraphics[scale=0.4]{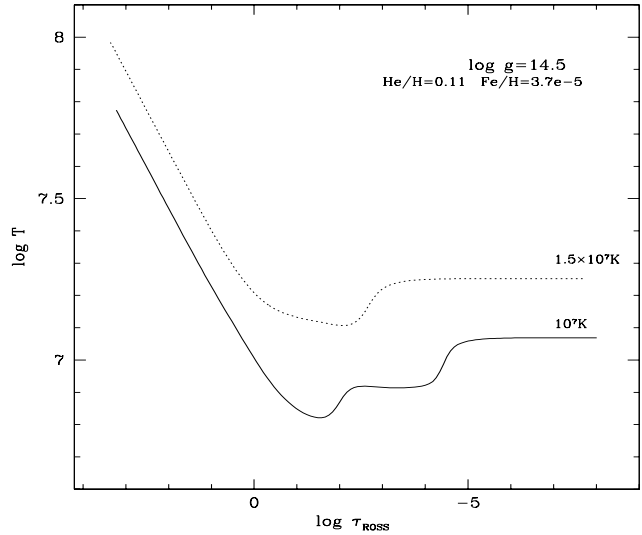}}}
\caption{\small{ Temperature structure of atmospheres with solar iron 
abundance and $\log g=14.5$, and of different $\te$; solid line -- model 
of $\te=10^7$ K, dotted line -- $1.5 \times 10^7$ K. }}
\label{f4}
\end{figure}

\begin{figure}[!h]
\resizebox{0.95\hsize}{7cm}{\rotatebox{0}{\includegraphics[scale=0.4]{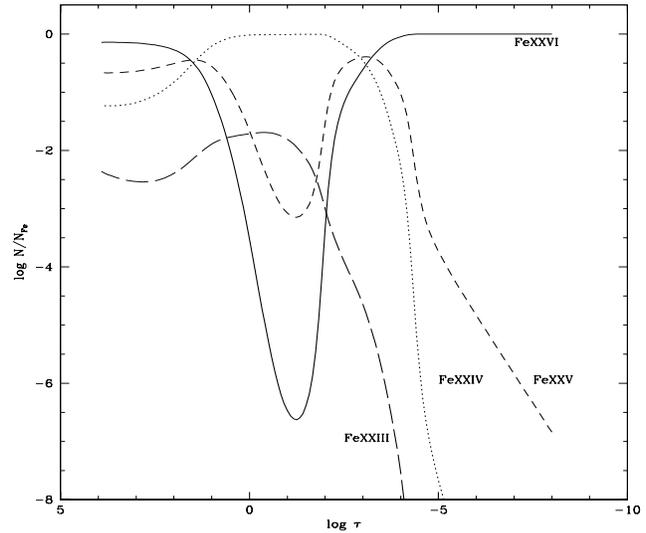}}}
\caption{\small{ Fractional abundances for various iron ions in the 
atmosphere with $\te =10^7$ K, $\log{g}=14.5$, and the solar iron
abundance.  }}
\label{f5}
\end{figure}

The run of temperature in our models can be quite complicated in some cases,
as is shown in Fig.~\ref{f4}. The relation $T( \tau_{Ross})$ displayed there
is implicitly determined by a very complicated distribution of various iron
ions and their continuum opacities, see also Fig.~\ref{f5}.   

\subsection{Model spectra}
Theoretical X-ray spectra of hot neutron stars differ significantly from
the blackbody spectra. In general, spectra of atmospheres in which Thomson
electron scattering opacity is present are harder than blackbody spectra
(Madej 1974; van Paradijs 1982). This effect is also displayed in Compton
scattering atmospheres, but then the frequency of the peak flux is lower.
It is located between the maxima of a blackbody with the effective 
temperature $\te$ and the spectrum of an atmosphere with coherent Thomson
scattering (Madej 1991a; Madej, Joss \& R\'o\.za\'nska 2004). 

This effect is caused by two factors. First, adding of scattering 
opacity to thermal absorption causes the layer of thermalisation
(where outgoing photons are created by thermal emission) move to 
deeper and hotter layers as compared to a purely absorbing atmosphere.
Consequently, the energy of the peak flux of the outgoing radiation increases.
Second, noncoherent Compton scattering of hot photons in cooler surface
layers removes (on average) part of the radiation energy. Due to this
effect, outgoing spectra are softer than those with Thomson scattering,
and the previous increase of the peak flux energy must be reduced. 

Such a situation can be reversed by simultaneous redistribution of radiation
flux caused by nongrey thermal absorption. In  hydrogen-helium atmospheres
of hot neutron stars both elements are almost completely ionised, and
then the only important thermal absorption is free-free absorption with
$\kappa_\nu \sim \nu^{-3}$. Absorption strongly decreases with increasing
energy, and it contributes to the additional increase of energy
at the peak. However, the situation can be more complicated in atmospheres
with nonzero iron abundance. 

\begin{figure}[!h]
\includegraphics[scale=0.4]{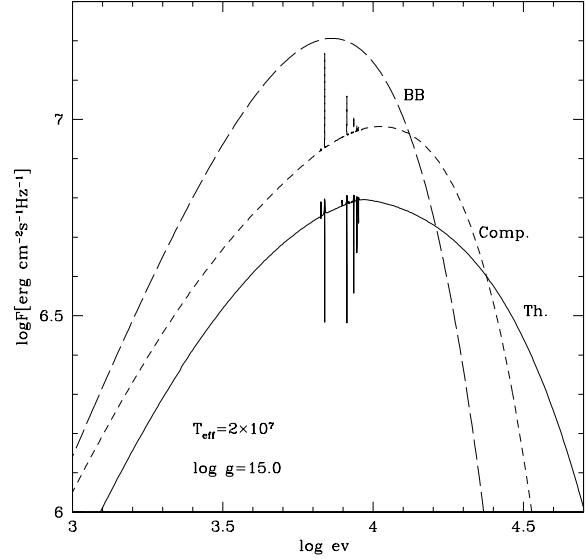}
\caption{\small{Comparison of outgoing spectra computed either with 
Thomson scattering or Compton scattering, and the blackbody spectrum.
All model spectra displayed here are computed with $\te=2\times 10^7$ K
and $\log{g}=14.5$. }}
\label{f6}
\end{figure}

As we can see in Fig.~\ref{f6}, energy of the peak flux of the
outgoing continuum radiation of an atmosphere with Compton scattering is
harder than in the case of the model atmosphere with Thomson scattering. 
In model atmospheres with iron, properties of outgoing radiation  and
its spectrum are determined by the temperature structure in the atmosphere,
which differs from $T(\tau)$ in H-He atmospheres. In iron rich models, 
atmospheres with Compton scattering are much hotter in opticaly thin layers
than in Thomson scattering case (Fig. \ref{f1}). This justifies
the result that spectra with Compton scattering can sometimes be
harder than spectra with Thomson scattering (Fig. \ref{f6}).

Comptonized spectra of hot neutron star atmospheres differ significantly
from blackbody spectra of the effective temperature $\te$, both in the case
of pure H-He and H-He-Fe chemical composition. The former model spectra
were taken from Madej, Joss \& R\'o\.za\'nska (2004). Differences between
model atmospheres and blackbody spectra depend on the $\te$, surface gravity
$\log g$ and the chemical composition.

\begin{figure}[!h]
\includegraphics[scale=0.4]{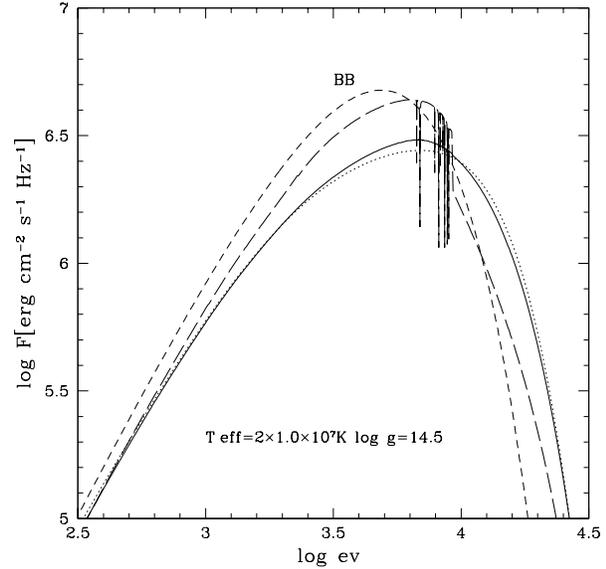}
\caption{\small{ Outgoing flux spectra of the atmosphere of
$\te=2\times 10^7$ K and $\log{g}=14.5$, and different iron abundances.
Long dashed line -- model with $N_{Fe}/N_H=3.7\times 10^{-3}$, 
solid line -- solar iron abundance $N_{Fe}/N_H=3.7\times 10^{-5}$, 
dotted line -- H-He atmosphere, dashed line -- blackbody spectrum.}}
\label{f7}
\end{figure}

\begin{figure}[!h]
\includegraphics[scale=0.4]{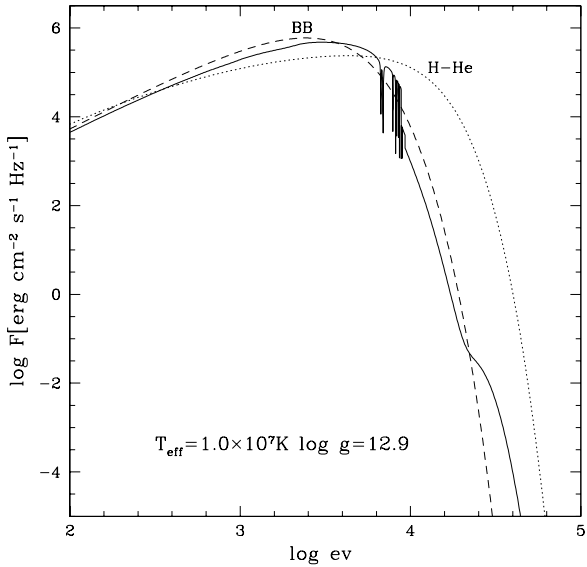}
\caption{\small{Flux spectra for $\te=10^7$K and $\log g=12.9$,
and different chemical composition. Solid line -- model with iron abundance
100 times solar value, dashed line -- the blackbody spectrum, 
dotted line -- model spectrum of pure H-He atmosphere.}}
\label{f8}
\end{figure}

All our spectra are harder than the blackbody of $\te$. Continuum 
outgoing spectra can be ordered in the following series, with increasing
hardeness: the blackbody, spectrum of an atmosphere with iron abundance 100
times higher than the solar value, spectra of atmospheres with solar iron
abundance ($N_{Fe}/N_H= 3.7\times 10^{-5}$), and spectra of pure H-He 
atmospheres (see Figs. \ref{f7} -- \ref{f8}).
This is because in an atmosphere with low or zero iron abundance, the ratio 
of thermal absorption to the total opacity coefficient is generally lower than
in an atmosphere of higher iron abundance. In a series of atmospheres with
the highest iron abundance $N_{Fe}/N_H= 3.7\times 10^{-3}$, monochromatic
thermal absorption generally increases due to the presence of strong bound-free
continuum absorption of lithium-like to hydrogen-like ions of iron, and that
absorption is strongest just beyond the corresponding spectral edge.
Moreover, ionised iron causes a very strong free-free absorption, which does
not produce spectral features and therefore remains invisible in outgoing
spectra.

\begin{figure}[!h]
\includegraphics[scale=0.4]{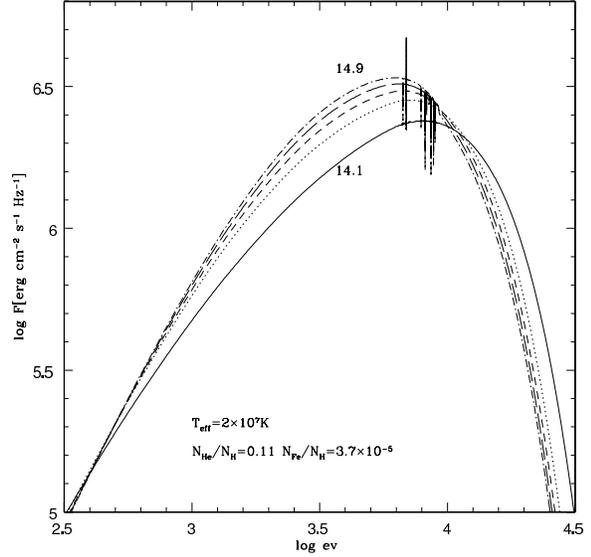}
\caption{\small{Comparison of theoretical spectra of atmospheres with
different $\log{g}$ and fixed $\te=2\times 10^7$ K and 
solar iron abundance. Solid line -- $\log{g}=14.1$ (it is very close to the
Eddington limit), dotted line -- 14.3, dashed line -- 14.5, long dashed line
-- 14.7, and the dot-dashed line -- 14.9.}}
\label{f9}
\end{figure}

Fig.~\ref{f9} presents the sequence of outgoing theoretical spectra
corresponding to the fixed $\te=2\times 10^7$ K and solar iron
abundance $N_{Fe}/N_H=3.7\times 10^{-5}$, and various $\log g$. We can see
how theoretical X-ray spectra evolve towards higher energies, while the
surface gravity $\log g$ decreases towards the critical gravity $g_{cr}$. 
 
Surface gravity $\log g $ significantly modifies the structure of an
atmosphere because it directly determines stratification of density $\rho$
through the equation of hydrostatic equilibrium. The spectrum of 
outgoing radiation emitted from the atmosphere of a neutron star of low 
surface gravity $\log g$ is much harder than spectra corresponding to
high $\log g$. Atmospheres of low $\log g$ are more extended and have lower
densities than the atmosphere of a neutron star with higher gravity.
Therefore, Compton scattering opacity gives a higher contribution to the total
opacity in the former cases. The spectrum of an atmosphere with high surface
gravity shows two prominent absorption edges, caused by hydrogen-like and
helium-like iron. We also see deep absorption lines belonging to the
fundamental series of both iron ions. 

We have included in our model atmosphere computations the series of lines
$1 ^1 S_0 - n ^1 P^0_1$ of helium-like iron, and the series
$1 ^2 S_{1/2} - n ^2 P_{1/2,3/2}^0$ of hydrogen-like iron. In both cases
the principal quantum number of the upper level $n=2, \ldots , 6$, and 
higher lines in both series were not included. Therefore we calculated 
profiles of a total of 10 lines. In this work we ignored the series of 
intercombination lines $1 ^1 S_0 - n ^3 P_1^0$ of helium-like iron, which
should be of noticeable strength in highly ionized iron.

All the above lines cause a blanketing effect on model atmospheres
computed, and therefore they influence the radiative equilibrium and
the run of temperature $T(\tau)$ in all models. Therefore, broadening of these
lines is of particular importance.  We have considered iron
lines broadened by natural (radiative), thermal (Doppler), and pressure
(Stark) broadenings, the latter considered following Griem (1974). The final
opacity profile of each of those lines was carefully
computed as the convolution of all of the three mechanisms (Madej 1989). 

\subsection{The ratio of color to effective temperatures}

Tables 1 and 2 present the ratios of color and effective temperatures for all
the computed models. In the case of X-ray spectra computed with iron of
solar abundance, the ratios of color and effective temperature do not differ
strongly from those of pure H-He X-ray spectra (Madej, Joss \& R\'o\.za\'nska
2004). In both series of models (H-He atmospheres and those with
solar iron abundance) we note the existence of local minima of relations
$T_c/\te $ vs. $\log g$ for low effective temperatures ($\te =1,1.5\times
10^7$ K). For higher effective temperatures ($\te =2\times 10^7$ or even
higher) the ratio $T_c/\te $ increases with decreasing surface gravity.

Table 1 shows that for low effective temperatures these ratios decrease 
with decreasing surface gravity from 1.35 (at $\log{g}=15.0$) to 1.28 
(at $\log{g}=14.3$). The local minimum of $T_c/\te $ is rather flat, and
its minimum value of 1.28 corresponds to a wide range of surface gravities,
ranging from $\log{g}=14.4$ to $\log{g}=13.8$. For lower $\log g$ the ratio
$T_c/\te $ increases from 1.29 (for $\log{g}=13.7$) to 1.62 (for 
$\log{g}=12.9$), the latter value corresponding to a model which is
very close to the critical surface gravity. The local minimum of 
$T_c/\te $ mentioned above separates regions on the $\log g$ axis where
the outgoing spectrum (and the energy of peak flux) is determined mostly
by the redistribution of radiation by highly nongray absorption and
effects of Compton scattering, respectively. 

In spectra of neutron stars of high effective temperatures, the ratios 
$T_c/\te $ increase with decreasing surface gravity $\log{g}$. This is 
because in very hot atmospheres the ratio of Compton scattering
opacity to the total opacity coefficient at some fixed frequency increases
and their spectra differ more and more from a blackbody. The highest value
of the ratio of color and effective temperature in our iron-rich models
approaches 1.77, and is obtained for the highest effective temperature
of our grid, $\te =3\times 10^7$ K, and $\log{g}=14.8$.
We obtain quite a large value of $T_c/\te =2.01$ also for $\log{g}=14.4$,
$\te =2.5\times 10^7$, and the solar abundance of iron. In the latter
model 96\% of the gravitational acceleration is compensated for by the
acceleration exerted by the radiation field.

For spectra of atmospheres with an iron abundance 100 times higher than the
solar value, the $T_c/\te $ ratios are lower than those for atmospheres with
lower iron abundance for the same $\te $ and $\log{g}$, see Table 2. This is
because in atmospheres with overabundant iron the importance of thermal 
absorption is higher than in atmospheres with solar iron abundance. In models
of low effective temperatures ($\te = 1\times 10^7$ K and $1.5\times 10^7$ K),
the ratio $T_c/\te $ decreases with decreasing surface
gravity $\log g$. For $\te =2\times 10^7$ K and  higher, these
ratios increase with decreasing $\log g $ due to increasing significance
of Compton scattering. 
\begin{figure}[!h]
\includegraphics[scale=0.4]{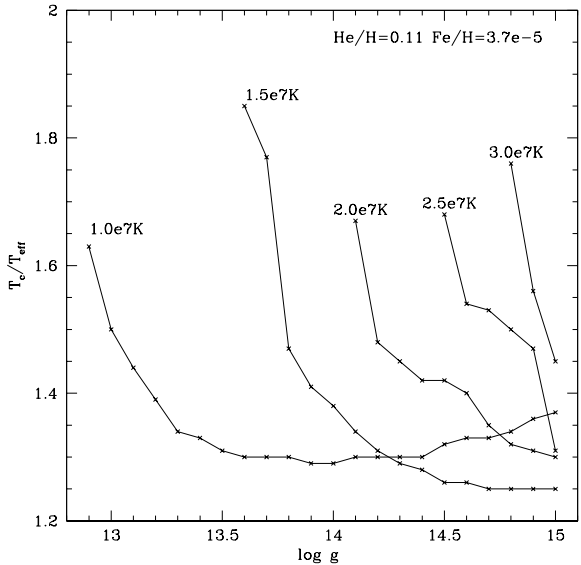}
\caption{\small{Run of ratios $T_c/\te$ vs. $\log g$ in models with
solar iron abundance}}
\label{f10}
\end{figure}
\begin{figure}[!h]
\includegraphics[scale=0.4]{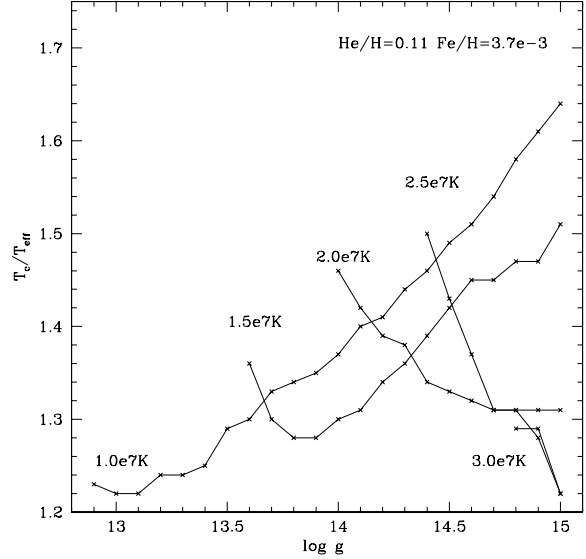}
\caption{\small{Run of ratios $T_c/\te$ vs. $\log g$ in models with
iron abundance 100 times higher than the solar value. }}
\label{f11}
\end{figure}

\section{Summary and discussion}

We have computed and presented two grids of model atmospheres and theoretical
spectra for hot neutron stars, which correspond to X-ray burst sources.
The assumed chemical compositions include hydrogen and helium in solar
proportion, $N_{He}/N_H = 0.11$, and iron with two different
abundances, $N_{Fe}/N_H = 3.7\times 10^{-5}$ and $3.7\times 10^{-3}$ 
(the solar abundance and 100 times solar). All spectra were calculated on a
large grid of effective temperatures $\te$ and surface gravities $\log{g}$.
We calculated two grids of models {bf with} various iron abundances for 5
effective temperatures:  1, 1.5, 2, 2.5, and 3$\times 10^7$ K, and with
many surface gravities ranging from $\log g=15$ (in cgs units) down to the
logarithm of critical gravity, $\log g_{cr}$, in steps of $\Delta \log g
= 0.1$.

We assumed an LTE equation of state for an ideal gas, both the radiative and
hydrostatic equilibrium, zero magnetic field, no neutron-star rotation,
and plane-parallel geometry for the model atmospheres. The equation of
radiative transfer includes free-free absorption from all ions, 
bound-free absorption from all ground levels of H, He, and Fe ions, and terms
describing Compton scattering by free electrons.
Bound-bound opacities (spectral lines) of helium-like and hydrogen-like iron 
are also included (10 lines of the fundamental series for both ions).
The presence of scattering terms implies that radiative transfer is a
non-LTE process. 

All model atmospheres computed in this work exhibit a temperature
inversion in the uppermost layers. This is caused by the heating effect of
Compton scattering on the cooler electron gas by hotter radiation from
deep thermalisation layers. The same heating effect has been found in
pure H-He model atmospheres (Madej, Joss \& R\'o\.za\'nska 2004). 

Our calculations show that all model spectra of X-ray bursters differ from
blackbody spectra at any given effective temperature $\te$. 
The spectral differences depend on the parameters of a particular
model atmosphere, i.e., $\te$, $\log g$, and chemical composition. 
The continuum spectra of iron-rich atmospheres that were presented in
this paper are always harder than a blackbody spectrum at a given $\te$.
Similar results were also presented by Madej, Joss \& R\'o\.za\'nska
(2004) for H-He atmospheres. Therefore, the color temperature $T_c$ 
determined from the peak flux frequency is always higher than the
effective temperature $\te$ in neutron-star atmospheres 
dominated by Compton scattering. The ratio $T_c/\te$ does not exceed
1.85 in our models. 

The theoretical spectra presented in this paper display two series of iron
spectral lines, belonging to helium-like and hydrogen-like iron (fundamental
series). Line profiles are computed in LTE, which means that the line source
function is the Planck function. Profiles of these lines are computed as the
numerical convolution of natural, thermal and pressure broadening, the
latter approximated on the basis of Griem (1976); see also Madej (1989). 
A few of the strongest absorption lines exhibit reverse emission in the line
cores, which was caused by the temperature inversion in the higher layers
of the model atmospheres. 

It is particularly interesting and new in our calculations that the peak
fluxes of spectra with Compton scattering sometimes exhibit higher 
color temperatures $T_c$ than Thomson scattering spectra computed
for the same $\te$ and $\log g$. In the spectra of pure H-He atmospheres
the situation is always different. In the latter case the peak fluxes of
spectra with Compton scattering at a given $\te$ are located 
at energies above the maximum for a blackbody spectrum but below
the maximum for a Thomson scattering spectrum.

This paper is an improvement and extention of the work by London,
Taam \& Howard (1986). These authors computed 17 models of X-ray burst 
sources, but only 3 models assumed nonzero iron abundance (with a solar
value of $N_{Fe}/N_H=10^{-5}$).
Therefore, a comparison between their models and ours is very
difficult. The $T_c/\te $ ratios obtained by London, Taam \& Howard
(1986) are larger than our values.
For example: for $\te =1.17\times 10^7$ K and $\log{g}=15.0$ we obtained
$T_c/\te =1.3$, while London, Taam \& Howard (1986) obtained 1.45; for
$\te =1.75\times 10^7$ K and $\log{g}=15.0$: 1.25 and 1.36, respectively;
and for $\te =1.25\times 10^7$ K and $\log{g}=14.0$: 1.32 and 1.46. 

Kuulkers et al. (2002) argued that if a burst aproaches the Eddington limit,
then the deviation of its spectrum from blackbody is the largest, 
$T_c/\te \sim 2$, and that higher values of $T_c/\te$ do not occur. 
We confirm the claim with our precise numerical calculations. 

We note that our extensive grids of models are an excellent tool for
fitting the observed spectra of X-ray bursts by use of, for example,
the {\sc xspec} software. Such spectral fitting is most reliable
when the atmosphere of the neutron star can be regarded as static, remaining
in both hydrostatic and radiative equilibrium. These conditions are
best fulfiled in the cooling phase of an X-ray burst and during the
quiescent intervals between bursts.

\begin{acknowledgements}

This work has been supported by the Polish Committee for
Scientific Research grant No. 1 P03D 001 26.

\end{acknowledgements}


\begin{table}[!h]
\caption{$T_c/T_{eff}$ ratios in models with $N_H/N_{He}=0.11$, and 
$N_{Fe}/N_{H}=10^{-5}$.}
  $$
\begin{array}{||c|c|c|c|c|c||} 
\hline \hline
\log{g}\diagdown T_{eff}&1\times 10^7&1.5\times 10^7&2\times 10^7&2.5\times
10^7&3\times 10^7\\ \hline\hline
15.0 & 1.37 & 1.25 & 1.30 & 1.31 & 1.45 \\ 
14.9 & 1.36 & 1.25 & 1.31 & 1.47 & 1.56\\
14.8 & 1.34 & 1.25 & 1.32 & 1.50 & 1.76\\
14.7 & 1.33 & 1.25 & 1.35 & 1.53 & -\\
14.6 & 1.33 & 1.26 & 1.40 & 1.54 & -\\
14.5 & 1.32 & 1.26 & 1.42 & 1.68 & -\\
14.4 & 1.30 & 1.28 & 1.42 & -&- \\
14.3 & 1.30 & 1.29 & 1.45 & -& -\\
14.2 & 1.30 & 1.31 & 1.48 & -& -\\
14.1 & 1.30 & 1.34 & 1.67 & -& -\\
14.0 & 1.29 & 1.38 & -    & -& -\\
13.9 & 1.29 & 1.41 & -    & -& -\\
13.8 & 1.30 & 1.47 & -    & -& -\\
13.7 & 1.30 & 1.77 & -    & -& -\\
13.6 & 1.30 & 1.85 & -    & -& -\\
13.5 & 1.31 & -    & -    & -& -\\
13.4 & 1.33 & -    & -    & -& -\\
13.3 & 1.34 & -    & -    & -&- \\
13.2 & 1.39 & -    & -    & -&- \\
13.1 & 1.44 & -    & -    & -& -\\
13.0 & 1.50 & -    & -    & -& -\\
12.9 & 1.63 & -    & -    & -&- \\ \hline\hline
\end{array}
   $$
\end{table}
\begin{table}[!t]
\caption{$T_c/T_{eff}$ ratios in models with  $N_H/N_{He}=0.11$, and 
$N_{Fe}/N_{H}=10^{-3}$.}
  $$
\begin{array}{||c|c|c|c|c|c||} 
\hline \hline
\log{g}\diagdown T_{eff}&1\times 10^7&1.5\times 10^7&2\times 10^7&2.5\times
10^7&3\times 10^7\\ \hline\hline
15.0 & 1.64 & 1.51 & 1.31 & 1.22 & 1.22 \\ 
14.9 & 1.61 & 1.47 & 1.31 & 1.28 & 1.29\\
14.8 & 1.58 & 1.47 & 1.31 & 1.31 & 1.29 \\ 
14.7 & 1.54 & 1.45 & 1.31 & 1.31 & -\\
14.6 & 1.51 & 1.45 & 1.32 & 1.37 & -\\
14.5 & 1.49 & 1.42 & 1.33 & 1.43 & -\\
14.4 & 1.46 & 1.39 & 1.34 & 1.50 &- \\
14.3 & 1.44 & 1.36 & 1.38 & -& -\\
14.2 & 1.41 & 1.34 & 1.39 & -& -\\
14.1 & 1.40 & 1.31 & 1.42 & -& -\\
14.0 & 1.37 & 1.30 & 1.46 & -& -\\
13.9 & 1.35 & 1.28 & -& -& -\\
13.8 & 1.34 & 1.28 & -& -& -\\
13.7 & 1.33 & 1.30 & -& -& -\\
13.6 & 1.30 & 1.36 & -& -& -\\
13.5 & 1.29 & -    & -& -& -\\
13.4 & 1.25 & -    & -& -& -\\
13.3 & 1.24 & -    & -& -&- \\
13.2 & 1.24 & -    & -& -&- \\
13.1 & 1.22 & -    & -& -& -\\
13.0 & 1.22 & -    & -& -& -\\
12.9 & 1.23 & -    & -& -&- \\ \hline\hline
\end{array}
   $$
\end{table}

\appendix
\section{The lower boundary condition}
At the lower boundary $\tau_{std}=\tau_{max}$ we assume that the radiation
field is thermalised at each frequency of the grid,
$\jnu(\tau_{max})=\bnu(\tau_{max})$. 

The second moment of the radiation field equals 
\begin{equation}
K_\nu=f_\nu \jnu =f_\nu \bnu^* \, ,
\end{equation}
where the Eddington factors $f_\nu=1/3$. The asterisk attached to the Planck
function indicates that its value strictly fulfils the condition of
radiative equilibrium. 

Differentiating of Eq. A.1 yields:
\begin{equation}
\frac{\partial}{\partial \tau_\nu}\left(f_\nu \jnu \right)=\frac{1}{3}\,
  \frac{\partial \bnu ^*}{\partial \tau_\nu}=\frac{1}{3\eta_\nu
  (\kappa+\sigma)_{std}\, \rho} \, \frac{\partial \bnu^*}{\partial T}\, 
  \frac{dT}{dz}
\end{equation}
where $d\tau_\nu=-\eta_\nu (\kappa+\sigma)_{std}\,\rho dz\, $, the ratio
$\eta_\nu=(\kappa_\nu+\sigma_\nu)/(\kappa+\sigma)_{std} \,$, and the
dimensionless absorption $\veps=\kappa_\nu /(\kappa_\nu+\sigma_\nu)$. 

The temperature gradient $dT/dz$, which is required in the above equation, can 
be determined from the bolometric flux: 
\begin{equation}
H=\frac{1}{3}\dnu \frac{1}{(\kappa_\nu +\sigma_\nu)\, \rho}\, \frac{\partial
  \bnu^*}{\partial T}\, \frac{dT}{dz} \, .
\end{equation}
Bolometric flux is given by: $H=\sigma_R \te^4 /4\pi \,$, where
$\sigma_R = 5.66961\times 10^{-5}$ erg cm$^{-2}$ s$^{-1}$ K$^{-4}$.  
Temperature gradient is
\begin{equation}
\frac{dT}{dz}=3H\times \left(\dnu \frac{1}{(\kappa_\nu +\sigma_\nu)\, \rho}
  \, \frac{\partial \bnu^*}{\partial T}\right)^{-1} \, ,
\end{equation}
therefore,
\begin{\row} 
\frac{\partial}{\partial \tau_\nu}\left(f_\nu \jnu 
\right)&=&\frac{H}{\eta_\nu(\kappa+\sigma)_{std}\, \rho} \, \frac{\partial 
\bnu^*}{\partial T} \\ \nonumber
&\times& \left(\dnu \frac{1}{(\kappa_\nu 
+\sigma_\nu)\, \rho}\, \frac{\partial \bnu^*}{\partial T}\right)^{-1} \, .
\end{\row}
We approximate $\partial \bnu^*/ \partial T$ in the above equation by a
linear Taylor series
\begin{equation}
\frac{\partial \bnu^*}{\partial T}=\frac{\partial \bnu}{\partial
  T}+\Delta T \, \frac{\partial^2 \bnu}{\partial T^2} \, ,
\end{equation}
and obtain the lower boundary condition in the following expanded form:
\begin{\row} 
\frac{\partial}{\partial \tau_\nu}\left(f_\nu \jnu 
\right)=\frac{H}{\eta_\nu  (\kappa+\sigma)_{std}\, \rho} \, 
\left(\frac{\partial \bnu}{\partial  T}+\Delta T \frac{\partial^2 
\bnu}{\partial T^2}\right) \\ \nonumber
 \times \left[\dnu \frac{1}{(\kappa_\nu 
+\sigma_\nu)\, \rho}\left(\frac{\partial \bnu}{\partial  T}+\Delta T 
\frac{\partial^2 \bnu}{\partial T^2}\right)\right]^{-1} \, . 
\end{\row}
Let us introduce new variables:
\begin{equation}
M_1=\dnu \frac{\partial \bnu}{\partial T} \, 
\frac{1}{\kappa_\nu+\sigma_\nu} \, ,
\end{equation}
\begin{equation}
M_2=\dnu \frac{\partial^2 \bnu}{\partial T^2} \, 
\frac{1}{\kappa_\nu+\sigma_\nu} \, .
\end{equation}
The lower boundary condition assumes the following form:
\begin{\row}
&&\frac{\partial}{\partial \tau_\nu}\left(f_\nu \jnu 
\right)=\frac{H\cdot M_1^{-1}}{\kappa_\nu+\sigma_\nu}\, 
\left(\frac{\partial \bnu}{\partial T} + \Delta T \, \frac{\partial^2 
\bnu}{\partial T^2} \right) \times \\ \nonumber 
&&\hskip2cm \times \, (M_1+M_2 \Delta T)^{-1}  \\ \nonumber
&&\approx \frac{H\cdot 
M_1^{-1}}{\kappa_\nu+\sigma_\nu}\, \frac{\partial \bnu}{\partial T} 
\times \left[1+\Delta T \left(\frac{\partial \bnu}{\partial T}\right)^{-1} 
\frac{\partial^2 \bnu}{\partial T^2} \, \frac{M_2}{M_1} \, \Delta T \right]
\end{\row}
Neglecting terms with $(\Delta T)^2$ we obtain:
\begin{\row}
\frac{\partial}{\partial \tau_\nu}\left(f_\nu \jnu 
\right)&=&\frac{H\cdot M_1^{-1}}{\kappa_\nu+\sigma_\nu}\, 
\frac{\partial \bnu}{\partial T} \times \\ \nonumber
&\times& \left[1+ \Delta T \left(\frac{\partial \bnu}{\partial 
T}\right)^{-1} \, \frac{\partial^2 \bnu}{\partial T^2} - \frac{M_2}{M_1} \, 
\Delta T \right] \, .
\end{\row} 
Temperature corrections $\Delta T$ are necessary to complete the above 
equation. These corrections can be obtained from the equation of radiative
equilibrium:
\begin{\row}
\label{equ:rownowaga}
0&=& \dnu \, \eta_\nu \veps (\jnu-\bnu^{\ast}) \\ \nonumber
&+&\dnu\, (1-\veps)\jnu \calka \Phi_1^{\ast}\naw \\ \nonumber
&-&\dnu\, \eta_\nu (1-\veps)\calka \Phi_2^{\ast}\naw \, \jp \, .
\end{\row}
Eq.~\ref{equ:rownowaga} should be linearized with respect to temperature $T$,
in order to replace variables with an asterisk:
\begin{\row}
&& T^*(\tau_\nu)=T(\tau_\nu)+\Delta T(\tau_\nu) \, , \\ 
&&\bnu^\ast(T) =\bnu(T)+\Delta T \left(\frac{\partial \bnu}{\partial 
T}\right)_\tau \, , \\&&\Phi_1^\ast\naw = \Phi_1\naw +\Delta T 
\left(\frac{\partial \Phi_1}{\partial T}\right)_\tau \, , \\
&&\Phi_2^\ast\naw = \Phi_2\naw +\Delta T \left(\frac{\partial 
\Phi_2}{\partial T}\right)_\tau \, .
\end{\row}
This step yields temperature corrections in the form:
\begin{\row}
\Delta T=\frac{\dnu\, \veps \eta_\nu\, (\jnu-\bnu)+L(\tau)}{\dnu\, \veps\eta_\nu 
\left(\partial \bnu / \partial T\right) -L'(\tau)} \, ,
\end{\row}
where functions $L(\tau)$ and $L'(\tau)$ are defined as:
\begin{\row}
L(\tau)&=&\dnu (\jnu-\bnu) \calka \Phi_1\naw \times \\ \nonumber
&&\hskip1.5cm \left[ \eta_\nu (1-\veps)
  -\eta_{\nu^\prime}
  (1-\varepsilon_{\nu^\prime})\left(\frac{\nu^\prime}{\nu}\right)\right]\, ,
\end{\row}
\begin{\row}
L'(\tau)&=&\dnu \left[\eta_\nu (1-\veps)\jnu \calka \frac{\partial
\Phi_1}{\partial T} \right. \\ \nonumber 
&& \hskip1.5cm \left. - \eta_\nu (1-\veps)\calka \jp \frac{\partial
\Phi_2}{\partial T}\right] \, .
\end{\row} 
Combining equations A.11 and A.17 we obtain the final form of the lower
boundary condition:
\begin{\row}
\frac{\partial}{\partial \tau_\nu}\left(f_\nu \jnu 
\right)&=&\frac{H\cdot M_1^{-1}}{\kappa_\nu+\sigma_\nu}\, 
\frac{\partial \bnu}{\partial T} \times \\ \nonumber
&\times& \left\{1+ \frac{\dnu \veps \eta_\nu(\jnu-\bnu)+L(\tau)}{\dnu 
\veps\eta_\nu \left(\partial \bnu / \partial T\right) -L'(\tau)} \times 
\right. \\ \nonumber
&&\left. \left[\left(\frac{\partial \bnu}{\partial T}\right)^{-1} \, 
\frac{\partial^2 \bnu}{\partial T^2} - \frac{M_2}{M_1} \right]\right\}\, .
\end{\row}
\end{document}